\documentclass[usenatbib]{mn2e}
\pdfoutput=1
\usepackage{amssymb}
\usepackage{amsfonts}
\usepackage{amsmath}
\usepackage[pdftex]{graphicx}
\usepackage{epsfig}
\usepackage{color}
\def\twid{\mathrel{\lower.1ex\hbox{$\sim$}}}
\def\gtwid{\mathrel{\raise.3ex\hbox{$>$\kern-.75em\lower1ex\hbox{$\sim$}}}}
\def\ltwid{\mathrel{\raise.3ex\hbox{$<$\kern-.75em\lower1ex\hbox{$\sim$}}}}
\def\\{\hfil\break}
\def\ie{{\it i.e.\ }}
\def\eg{{\it e.g.\ }}
\def\etal{{\it et al.\ }}
\def\hmpc{$h^{-1}$Mpc}
\def\hkpc{$h^{-1}$kpc}

\def\kmSec{km s$^{-1}$}
\def\hMsun{$h^{-1}M_{\sun}$}
\newcommand{\degree}{{\rm o}}

\newcommand{\be}{\begin{equation}}
\newcommand{\ee}{\end{equation}}
\newcommand{\bea}{\begin{eqnarray}}
\newcommand{\eea}{\end{eqnarray}}

\begin{document}

\title[Minimum variance velocity moments]{Testing the minimum variance method for estimating large-scale
velocity moments}
\author[Agarwal \& Feldman \& Watkins]{
Shankar Agarwal$^{1, \star}$\,\& Hume A. Feldman$^{1, \dagger}$\,\& Richard Watkins$^{2,\ddagger}$\\
$^1$Department of Physics \& Astronomy, University of Kansas, Lawrence, KS 66045, USA.\\
$^2$Department of Physics, Willamette University, Salem, OR 97301, USA.\\
emails: $^{\star}$sagarwal@ku.edu; $^{\dagger}$feldman@ku.edu; $^{\ddagger}$rwatkins@willamette.edu}
\date{} 

\maketitle

\begin{abstract}

The estimation and analysis of large-scale bulk flow moments of peculiar velocity surveys is complicated by non-spherical survey geometry, the non-uniform sampling of the matter velocity field by the survey objects and the typically large measurement errors of the measured line-of-sight velocities. Previously, we have developed an optimal `minimum variance' (MV) weighting scheme for using peculiar velocity data to estimate bulk flow moments for idealized, dense and isotropic surveys with Gaussian radial distributions, that avoids many of these complications. These moments are designed to be easy to interpret and are comparable between surveys. In this paper, we test the robustness of our MV estimators using numerical simulations. Using MV weights, we estimate the bulk flow moments for various mock catalogues extracted from the LasDamas and the Horizon Run numerical simulations and compare these estimates to the moments calculated directly from the simulation boxes. We show that the MV estimators are unbiased and negligibly affected by non-linear flows.

\end{abstract}




\section{Introduction}
\label{sec:INTRO}

Peculiar velocities are a sensitive probe of the underlying large-scale matter density fluctuations in our Universe.   In particular, large, all-sky surveys of the peculiar velocities of galaxies or clusters of galaxies can provide important constraints on cosmological parameters.   However, studies of peculiar velocities suffer from several drawbacks, including (i) the presence of small-scale, non-linear flows, such as infall into clusters, can potentially bias analyses which typically rely on linear theory, (ii) sparse, non-uniform sampling of the peculiar velocity field can lead to aliasing of small-scale power on to large scales and bias due to heavier sampling of dense regions, (iii) large measurement uncertainties of individual peculiar velocity measurements, particularly for distant galaxies or clusters, make it necessary to work with large surveys in order to extract meaningful constraints.

These difficulties have often been addressed by calculating statistics from peculiar velocity surveys that are designed to primarily reflect large-scale flows which are well described by linear theory.   The most common statistic used is the bulk flow, which represents the average motion of the objects in a survey.   The bulk flow statistic has been investigated extensively by many groups (\citealt{DreFab90,Kai91,FelWat94,JafKai95,StrCenOstLauPos95,WatFel95,HudSmiLuc99,HudSmiLuc04,daCBerAlo00b,ParTug04}; \citealt*{SarFelWat07}; \citealt{KasAtrKoc08,KasAtrKoc10}; \citealt*{MaGorFel11}; \citealt{MacFelFer11}; \citealt*{NusBraDav11}; \citealt{NusDav11,AbaFel12,TurHudFel12}).  However, bulk flow estimates can be difficult to interpret since how they sample the peculiar velocity field depends strongly on the characteristics of the particular survey being considered.   In addition, results from bulk flow analyses have often been controversial, highlighting the importance of developing a robust bulk flow statistic that is easy to interpret and that can be compared between surveys with different geometries.

In \citealt*{WatFelHud09} (hereafter Paper I) and \citealt*{FelWatHud10} (hereafter Paper II), we developed the `minimum variance' (MV) moments that were designed to estimate the bulk flow of a volume of a given scale rather than a particular peculiar velocity survey.   We stress that the MV moments do not represent the bulk motion of the galaxies in a survey,  rather they are estimates of the bulk motion of a given volume of space.   The MV algorithm was designed to make a clean estimate of the large-scale bulk flow as a function of scale using the available peculiar velocity data. Essentially, each velocity datum in a real survey is weighted in a way that minimizes the variance of the difference between the MV-weighted bulk flow of the real survey and an idealized survey bulk flow, on a characteristic scale R. The MV analysis suggested bulk flow velocities well in excess of expectations from a $\Lambda$ cold dark matter ($\Lambda$CDM) model with 7-year {\it{Wilkinson Microwave Anisotropy Probe}} (WMAP7; \citealt{wmap7}) central parameters.

Indeed there are a few recent observations that suggest that the standard model may be incomplete: large-scale anomalies found in the maps of temperature anisotropies in the cosmic microwave background (CMB; \citealt{CopHutSch10,SarHutCop11,WMAPanom10});
a recent estimate \citep{LeeKom10} of the occurrence of high-velocity merging systems such as the bullet cluster is unlikely at a $\sim\!\!6\sigma$ level; 
large excess of power in the statistical clustering of luminous red galaxies (LRG) in the photometric Sloan Digital Sky Survey (SDSS) galaxy sample \citep*{ThoAbdLah11};
\citet*{KovBenItz10} find a unique direction in the CMB sky determined by anomalous mean temperature ring profiles, also centred about the direction of the flow detected above;
larger than expected cross-correlation between samples of galaxies and lensing of the CMB \citep{HoHirPad08,HirHoPad08};
Type Ia Supernovae (SNIa) seem to be brighter than expected at high redshift \citep{Kowalski08}; 
small voids ($\sim10$ Mpc) are observed to be much emptier than predicted \citep{GotLokKly03}; 
observations indicate denser high concentration cluster haloes than the shallow low concentration and density profile predictions \citep{deBlok05,Gentile05}.

In this paper, we use N-body simulations to investigate the robustness of our MV scheme for estimating the bulk flow moments of the velocity field, over a volume of a particular scale, R.   First we extract a mock catalogue (described in Sec.~\ref{sec:mocks}) from N-body simulations.   Given this mock catalogue, we use our MV algorithm (described in Sec.~\ref{sec:MV}) to estimate the bulk flow moments \{$u_x, u_y, u_z$\} of the velocity field over a volume of a particular scale.   Then we position ourselves in the N-body simulation box at the location of the centre of the mock catalogue, and calculate the Gaussian-weighted moments \{$V_x, V_y, V_z$\} by averaging the velocities of all the galaxies in the simulation box; each galaxy being weighted by a Gaussian radial distribution function $f(r)=e^{-r^2/2R^2}$.   Note that a large number of particles in the simulation box is preferable to accurately calculate the Gaussian moments of the velocity field.   Finally, we compare the MV-weighted moments \{$u_x, u_y, u_z$\} with the Gaussian-weighted moments \{$V_x, V_y, V_z$\} in Sec.~\ref{sec:BF}. A close match between the two would indicate that the MV scheme accurately estimates the Gaussian bulk flow on scale R.

It is worth mentioning here the reason for our choice of a Gaussian profile $f(r)$ over, for example, a Tophat filter in developing the MV formalism.   A Tophat filter gets contribution from small scales. As such, bulk flow calculated using a Tophat filter can be compared with theoretical expectations $\it{only}$ if the observed velocity field is reasonably dense and uniform, so that the small-scale systematics average out.   However, observations typically are sparse and non-uniform with large uncertainties.   This leads to aliasing of small-scale power on to large scales, making comparison with theory difficult.  A Gaussian filter, on the other hand, gets very little contribution from small scales and isolates the small-scale effects present in real surveys, thereby making comparison with theoretical predictions meaningful.   Our MV method is specifically designed to convert the observed velocity field into a Gaussian field on a user-specified scale $R$.

In Sec.~\ref{sec:MV} we review the MV formalism. In Sec.~\ref{sec:mocks} we describe the simulations we use and surveys we model to extract the mock catalogues. In sec.~\ref{sec:BF} we compare the MV-weighted bulk flow moments with the Gaussian-weighted moments. We discuss our results and conclude in Sec.~\ref{sec:Conclude}.

\section{The Minimum Variance Method}
\label{sec:MV}

Individual radial peculiar velocity measurements are plagued by large uncertainties and contributions from small-scale, non-linear processes which are difficult to model theoretically.   Both of these problems can be greatly reduced if instead of considering individual velocities an average velocity over a sample, commonly called the bulk flow, is worked with.   The three components of the bulk flow $u_i$ can be written as weighted averages of the measured radial peculiar velocities of a survey,
\begin{equation}
u_i= \sum_n w_{i,n} S_n ,
\label{eq:ui}
\end{equation}
where $S_n$ is the radial peculiar velocity of the $n$th galaxy of a survey, and $w_{i,n}$ is the weight assigned to this velocity in the calculation of $u_i$.   Throughout this paper, subscripts $i, j$ and $k$ run over the three components of the bulk flow, while subscripts $m$ and $n$ run over the galaxies. By far the most common weighting scheme used in studies of the bulk flow, which we will call the maximum likelihood estimate (MLE) method, is obtained from a maximum likelihood analysis introduced by \citet{Kai88}.   By modelling galaxy motions as being due to a uniform flow and assuming Gaussian-distributed measurement uncertainties, the likelihood function 
\begin{equation}
 L[u_i|\{S_n,\sigma_n,\sigma_*\}]= \prod_n {1\over \sqrt{\sigma_n^2 + \sigma_*^2}}\exp\left( -{1\over 2} {(S_n - \hat r_{n,i}}u_i)^2\over \sigma_n^2 + \sigma_*^2\right).
 \end{equation}
is obtained,  where ${\bf\hat r}_n$ is the unit position vector of the $n$th galaxy, $\sigma_n$ is the measurement uncertainty of the $n$th galaxy and $\sigma_*$ is a 1D velocity dispersion accounting for smaller scale motions. Maximizing this likelihood gives a bulk flow estimate of the form of Eq.~\ref{eq:ui}, with weights 
\begin{equation}
\label{eq:mle}
w_{i,n} = \sum_{j=1}^3 A_{ij}^{-1}{\hat r_{n,j}\over \sigma_n^2 + \sigma_*^2} ,
\end{equation}
where
\begin{equation}
A_{ij}= \sum_n {\hat r_{n,i}\hat r_{n,j}\over \sigma_n^2 + \sigma_*^2} .
\end{equation}
These weights play the dual roles of accounting for geometrical factors, \eg picking out the $x$ component of  velocities in a calculation of $u_x$, and down-weighting velocities with large uncertainties.   However, the fact that velocity uncertainties are typically proportional to distance, together with the sparseness of velocity catalogues at their outer edges, means that nearby objects are greatly emphasized in calculations of the MLE bulk flow.   Indeed, studies of the window functions of these moments (Paper I) have shown that MLE bulk flow moments of a survey are typically sensitive to flows on scales much smaller than the survey's physical diameter, thus complicating their interpretation.

In Paper I, we introduced an alternative to the MLE weights that yield bulk flow moments that are much easier to interpret.    First, we imagine an idealized survey containing radial velocities that well sample the velocity field in a region.   This survey consists of a large number of objects, all with zero measurement uncertainty.   For simplicity, the radial distribution of this idealized survey is taken to be a Gaussian profile of the form $f(r) \propto e^{-r^2/2R^2}$, where $R$ gives a measure of the depth of the survey.   This idealized survey has easily interpretable bulk flow components $U_i$ that are not affected by small-scale aliasing and which reflect the motion of a well-defined volume.   Note that the difference between $U_i$ and $V_i$ (see Sec.~\ref{sec:INTRO} for the definition of $V_i$) is that $U_i$ is calculated from an ideal (dense and isotropic) survey, while $V_i$ is based on the galaxy distribution obtained from N-body simulations. In the limit that the simulations are dense enough, $V_i$ will converge towards $U_i$.

Our goal is to construct estimators for the idealized survey bulk flow components $U_i$, out of the measured radial peculiar velocities $S_n$ and positions ${\bf r}_n$ contained in a real survey.   We assume that $S_n$ can be expressed as  $S_n= v_n+\delta_n$, where $v_n$ is the radial component of the linear peculiar velocity field at the location of the object and $\delta_n$ accounts for the measurement noise as well as any non-linear flow, \eg infall into a cluster.   In order to calculate the weights to use for the bulk flow estimators, we minimize the variance $\langle (u_i-U_i)^2\rangle$, where the average is over different realizations of a particular matter power spectrum.   Expanding this expression using Eq.~\ref{eq:ui} for the bulk flow estimate, we obtain

\begin{eqnarray}
\label{eq:variance}
\langle (u_i-U_i)^2\rangle &=& \sum_{m,n} w_{i,m}w_{i,n}\langle S_mS_n\rangle + \langle U_i^2\rangle  \\ \nonumber
&&- 2\sum_n w_{i,n}\langle  U_iv_n\rangle ,
\end{eqnarray}
where we have used the fact that the measurement error included in $S_n$ is uncorrelated with the bulk flow $U_i$.   

Before we minimize this expression with respect to the weights $w_{i,n}$, we impose the following constraint introduced in Paper II.   Suppose that the velocity field were a pure bulk flow, so that $S_n= \sum_i U_ig_i({\bf r}_n) + \delta_n$, where $U_i$ are the three bulk moments \{$U_x, U_y, U_z$\}; $g_i({\bf r}_n)$ are the direction cosines of the $n$th galaxy \{$\hat r_{n,x}, \hat r_{n,y}, \hat r_{n,z}$\} and $\delta_n$ is the noise due to measurement error.   We ask that the estimators $u_i$ give the correct amplitude for the flow on average (over different realizations of the universe), namely that $\langle u_i\rangle= U_i$.   Plugging the expression for $S_n$ into Eq.~\ref{eq:ui} give the constraint that
\begin{equation}
\label{eq:constraint}
\sum_n w_{i,n}g_j({\bf r}_n)= \delta_{ij},
\end{equation}

$\delta_{ij}$ being the Kronecker delta. This set of three constraints is implemented using Lagrange multipliers, so that we derive the desired weights by taking a derivative of the expression
\begin{eqnarray}
&& \sum_{m,n} w_{i,m}w_{i,n}\langle S_mS_n\rangle + \langle U_i^2\rangle 
- 2\sum_n w_{i,n}\langle  U_iv_n\rangle  \\ \nonumber
&&+\sum_{j=1}^3\lambda_{ij}\left(\sum_n w_{i,n}g_j({\bf r}_n)- \delta_{ij}\right)
\end{eqnarray}
with respect to $w_{i,n}$ and setting the resulting expression equal to zero.   Solving for the weights then gives
\begin{equation}
w_{i,n} = \sum_m G^{-1}_{mn}\left( \langle S_mU_i\rangle - {1\over 2}\sum_{j=1}^3 \lambda_{ij}g_j({\bf r}_m)\right) ,
\label{eq:weight}
\end{equation}
where $G$ is the covariance matrix of the individual measured velocities, $G_{mn}= \langle S_mS_n\rangle$.   The Lagrange multipliers can be found by plugging Eq.~\ref{eq:weight} into Eq.~\ref{eq:constraint} and solving for $\lambda_{ij}$, 
\begin{equation}
\lambda_{ij}= \sum_{k=1}^3\left[ M^{-1}_{ik}\left(\sum_{m,n}G^{-1}_{mn}\langle S_mU_k\rangle g_j({\bf r}_n)- \delta_{jk}\right)\right],
\end{equation}
where the matrix $M$ is given by 
\begin{equation}
M_{ij} = {1\over 2}\sum_{m,n} G^{-1}_{mn}g_i({\bf r}_n)g_j({\bf r}_m).
\end{equation}

In linear theory, the correlation $\langle S_mU_i\rangle$ and the covariance matrix $G$ that appear in our expression for $w_{i,n}$ can be calculated for a given matter power spectrum P(k) (for details see Paper II):

\begin{eqnarray}
\label{eq:qi}
\langle S_mU_i\rangle &=& \sum_{n^\prime=1}^{N^\prime} w^\prime_{i,n^\prime} \langle {S_mv_{n^\prime}}\rangle\\
&=& \sum_{n^\prime=1}^{N^\prime} w^\prime_{i,n^\prime} {H_0^2\Omega_{\rm m}^{1.1}\over{2\pi^2}}\int   dk\  P(k)f_{mn^\prime}(k), \nonumber
\end{eqnarray}

where
\begin{eqnarray}
w^\prime_{i,n^\prime} =  \sum_{j=1}^3 A_{ij}^{-1} {{{\hat r}_{n^\prime,j}^\prime} \over N^\prime} \nonumber
\end{eqnarray}

are the weights of an ideal, isotropic survey consisting of $N^\prime$ exact radial velocities $v_{n^\prime}$ measured at randomly selected positions ${\bf r}_{n^\prime}^\prime$ with

\begin{eqnarray}
A_{ij}= \sum_{n^\prime=1}^{N^\prime} {{\hat r}_{n^\prime,i}^\prime {\hat r}_{n^\prime,j}^\prime \over N^\prime},\nonumber
\end{eqnarray}

\begin{eqnarray}
 \label{eq:gmn}
G_{mn} &=& {H_0^2\Omega_{\rm m}^{1.1}\over{2\pi^2}}\int   dk\  P(k)f_{mn}(k)
+ \delta_{mn}(\sigma_*^2 + \sigma_n^2)\\ \nonumber
&=&\langle {\bf \hat r}_n\cdot {\bf v}({\bf r}_n)\ \   {\bf\hat r}_m\cdot {\bf v}({\bf r}_m)\rangle
+ \delta_{mn}(\sigma_*^2 + \sigma_n^2),
\end{eqnarray}

where $f_{mn}(k)$ is the angle-averaged window function:
\begin{eqnarray}
 \label{eq:fmn}
 f_{mn}(k) &=& \int {d^2{\hat k}\over 4\pi} \left( {\bf \hat r}_m\cdot {\bf \hat k} \right)\left( {\bf \hat r}_n\cdot {\bf \hat k} \right) \\ \nonumber
 &&\times\exp \left(ik\ {\bf \hat k}\cdot ({\bf r}_m - {\bf r}_n)\right).
\end{eqnarray}

Thus, given a peculiar velocity survey and a power spectrum model $P(k)$ we can calculate the optimum weights $w_{i,n}$ (see Eq.~\ref{eq:weight}) for estimating the MV moments (see Eq.~\ref{eq:ui}). We use the power spectrum model given by \cite{EisHu98} with WMAP7 \citep{wmap7} central parameters. Using the optimum weights $w_{i,n}$ from Eq.~\ref{eq:weight}, the angle-averaged tensor window function $W^2_{ij}(k)$ can be constructed (for details see Paper II) as
\begin{eqnarray}
 \label{eq:Wij}
 W^2_{ij}(k) &=& \sum_{m,n} w_{i,m} w_{j,n} f_{mn}(k).
\end{eqnarray}

The diagonal elements $W^2_{ii}$ are the window functions of the bulk flow components $u_i$. Given a velocity survey, $W^2_{ij}$ estimated using the MV weights are the closest approximation to the ideal window functions. See Paper I for the MV-estimated window functions of the bulk flow components for a range of surveys.

\begin{figure}
     \includegraphics[width= \columnwidth]{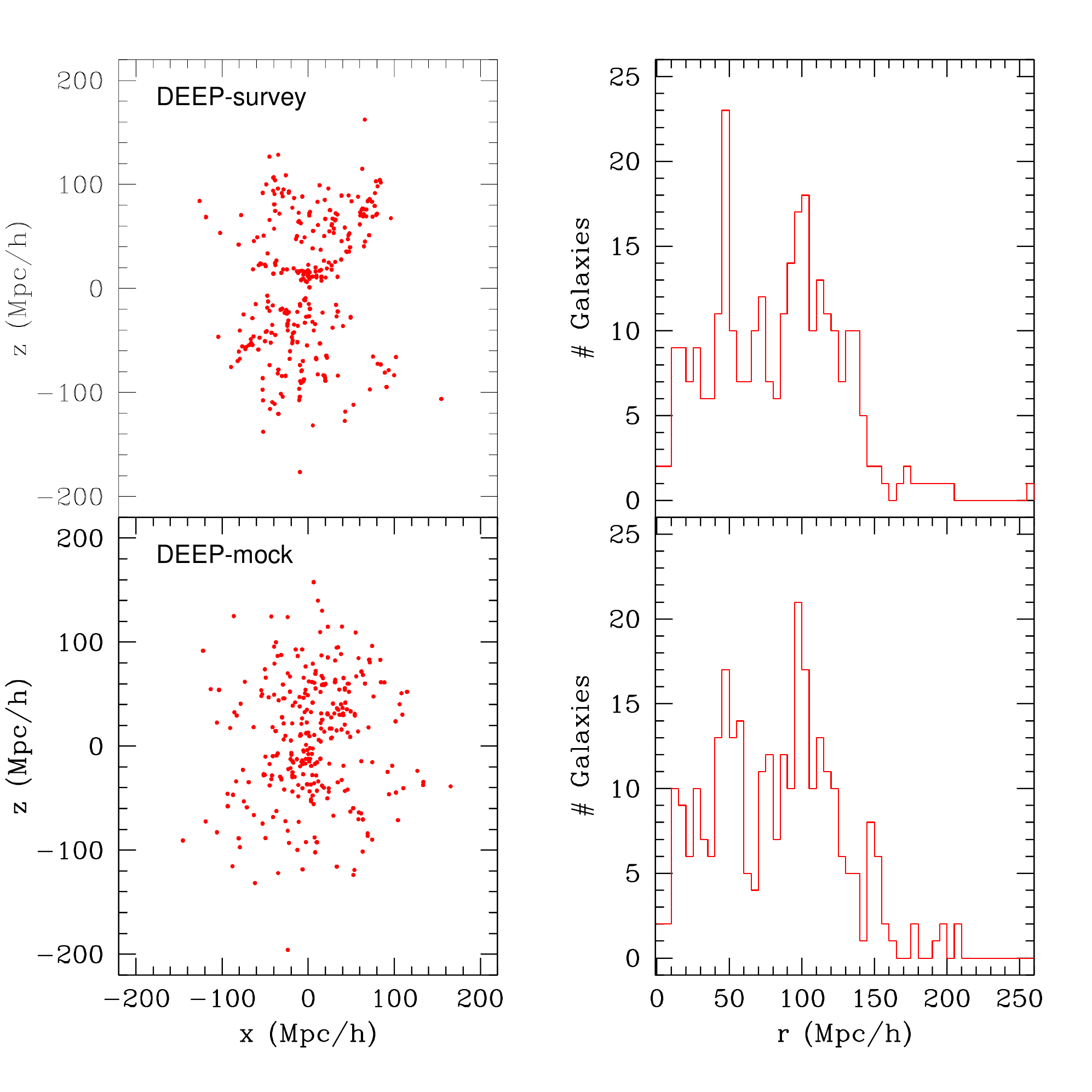}
        \caption{\small{Top row: DEEP catalogue (left) and its radial distribution (right). Bottom row: DEEP mock catalogue (left) and its radial distribution (right).
        }}
    \label{fig:DEEP}
\end{figure}

\section{Mock Catalogues}
\label{sec:mocks}

\subsection{N-body simulations}
\label{sec:SIM}

To check the robustness of our MV formalism, we calculated the bulk flow moments directly from numerical simulations. The N-body simulations we use in our analysis are (i) the Large Suite of Dark Matter Simulations (LasDamas; hereafter LD; \citealt{LasDamas}; McBride \etal 2011, in prep\footnote{http://lss.phy.vanderbilt.edu/lasdamas/download.html}) and (ii) Horizon Run (hereafter HR; \citealt{HR09}). These are designed to model the SDSS observations. The LD (HR) simulation parameters are $\Omega_{\rm m}=0.25\ (0.26)$, $\Omega_{\rm b}=0.04\ (0.044)$, $\Omega_\Lambda=0.75\ (0.74)$, $h=0.7\ (0.72)$, $\sigma_8=0.8\ (0.794)$, $n_{\rm s}=1.0\ (0.96)$ and $L_{\rm Box}=1\ (6.592)h^{-1} \textrm{Gpc}$ for the matter, baryonic and cosmological constant normalized densities, the Hubble parameter, the amplitude of matter density fluctuations, the primordial scalar spectral index and the simulation box size, respectively. The HR simulation samples the density field at $z=0$ and identifies galaxies using subhalos \citep*{KimParCho08}. The LD simulations, a suite of 41 independent realizations of dark matter N-body simulations named \textit{Carmen}, have information at $z=0.13$. Using the Ntropy framework \citep{GarConMcB07}, bound groups of dark matter particles (halos) are identified with a parallel friends-of-friends (FOF) code \citep{FOF}. The cosmological parameters and the design specifications of the LD-\textit{Carmen} and HR simulations are listed in Table~\ref{tab:parameters}.

\begin{figure}
     \includegraphics[width= \columnwidth]{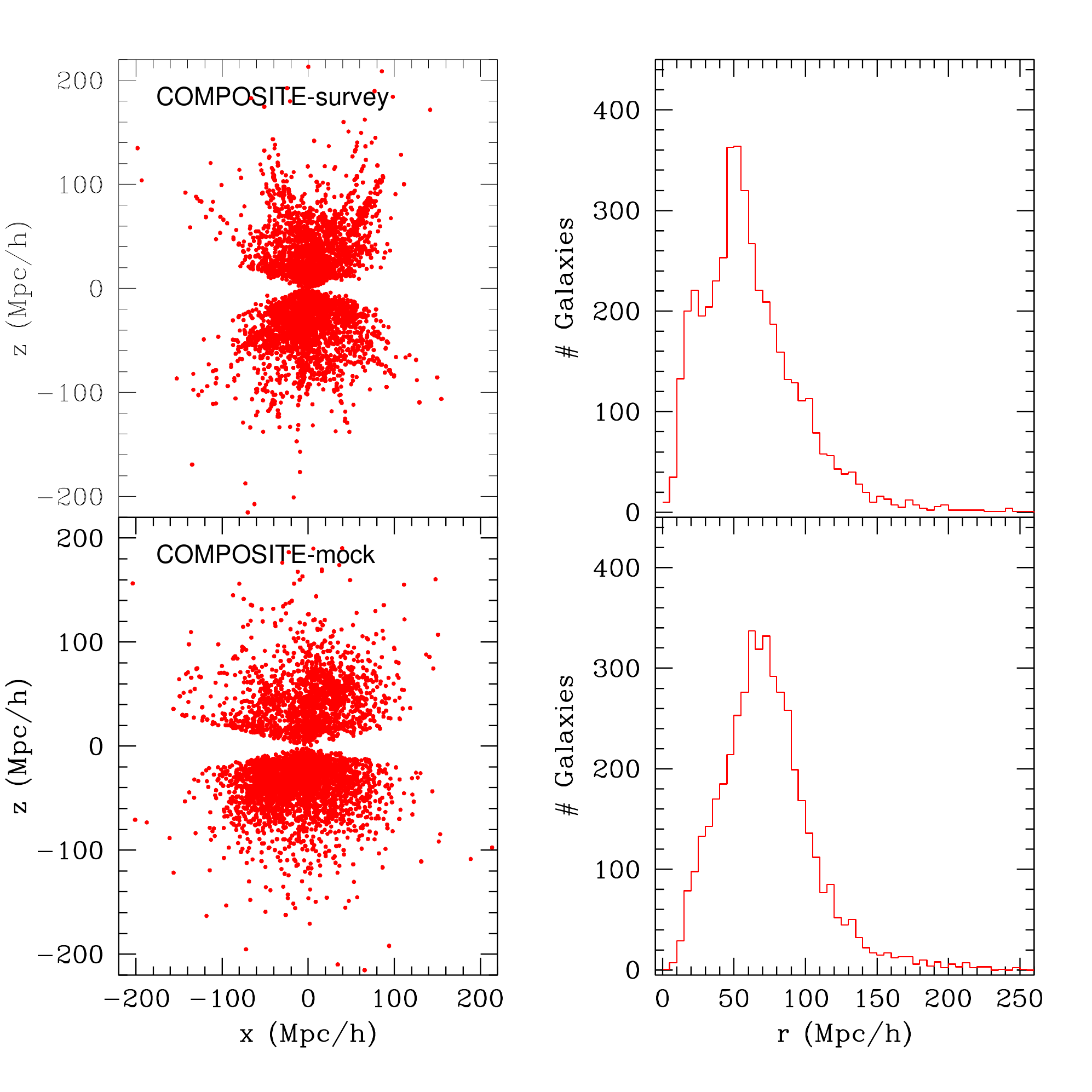}
        \caption{\small{Top row: COMPOSITE catalogue (left) and its radial distribution (right). Bottom row: COMPOSITE mock catalogue (left) and its radial distribution (right). The mock does not have as many close by objects as there are in the COMPOSITE catalogue.
        }}
    \label{fig:COMPOSITE}
\end{figure}

\begin{table*}
\caption{The cosmological parameters and the design specifications of the LD-\textit{Carmen} and HR simulations.}
\begin{tabular}{lll}
\hline
 & \multicolumn{1}{c}{LD-\textit{Carmen}} &  \multicolumn{1}{c}{HR} \\ \hline \\
 \multicolumn{1}{c}{Cosmological parameters} &  & \\\\
Matter density,						$\Omega_{\rm m}$				& 0.25 		& 0.26		\\
Cosmological constant density,		$\Omega_\Lambda$				& 0.75 		& 0.74		\\
Baryon density,						$\Omega_{\rm b}$				& 0.04 		& 0.044		\\
Hubble parameter,					$h$ (100 km s$^{-1}$ Mpc$^{-1}$)	& 0.7 		& 0.72		\\
Amplitude of matter density fluctuations,	$\sigma_8$					& 0.8			& 0.794		\\
Primordial scalar spectral index,		$n_{\rm s}$					& 1.0		 	& 0.96		\\\\
 \multicolumn{1}{c}{Simulation design parameters} & & \\\\  
 Simulation box size on a side (\hmpc)								& 1000		& 6592		\\
 Number of CDM particles											& 1120$^3$ 	& 4120$^3$	\\
 Initial redshift, $z$												& 49			& 23			\\
 Particle mass, $m_{\rm p}$ ($10^{10}$ \hMsun)						& 4.938		& 29.6		\\
 Gravitational force softening length, $f_\epsilon$ (\hkpc)					& 53			& 160		\\ \hline
   \end{tabular}
\label{tab:parameters}
\end{table*}

The LD-\textit{Carmen} data we use consists of 41 independent realizations, each in a $1h^{-1}$ Gpc box with the same initial power spectrum, but a different random seed. We extract 100 mock catalogues from each of the 41 LD boxes, for a total of 4100 mocks. The mock centres are randomly chosen inside the box. The mocks are extracted in a way that they come as close as possible to the radial distribution of real catalogues. The HR simulation is a single realization in a much bigger $6.592h^{-1} \textrm{Gpc}$ box. As such, we extract 5000 randomly distributed mocks.

\begin{figure*}
  \begin{flushleft}
   \centering
    \begin{minipage}[c]{1.0\textwidth}
      \centering
      \includegraphics[scale=0.33]{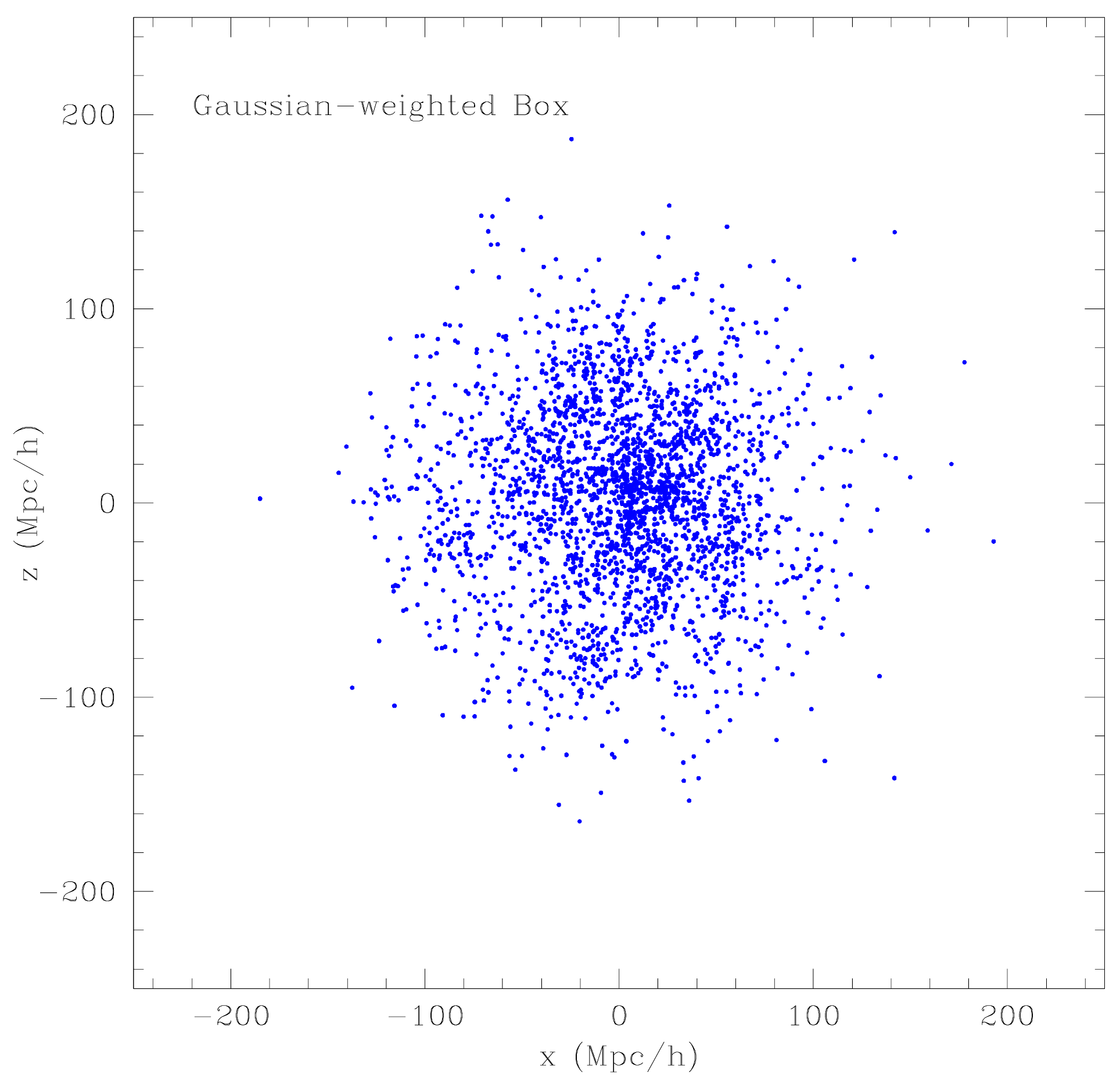}
      \includegraphics[scale=0.33]{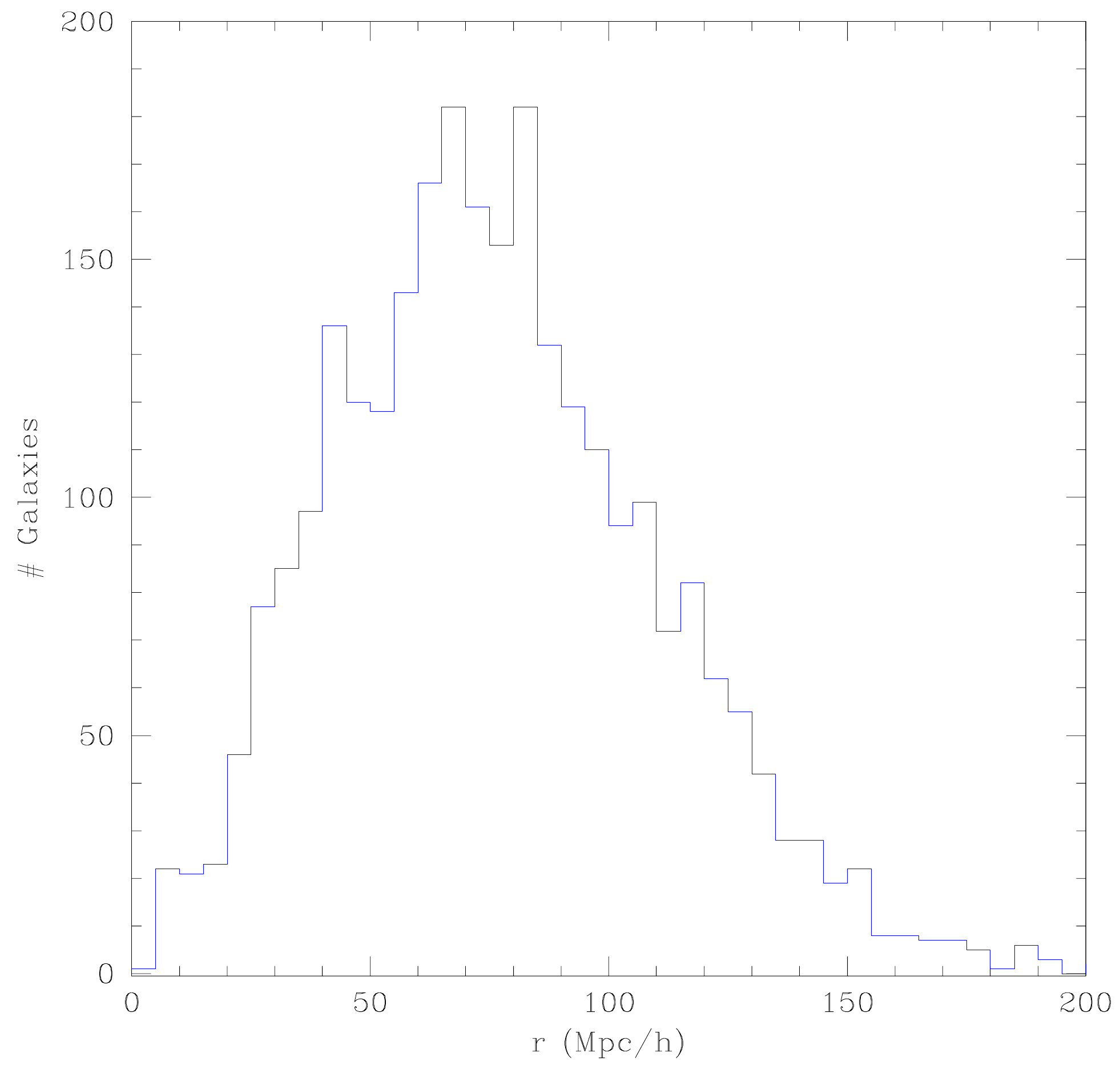}
    \end{minipage}
    \caption{\small{The left-hand panel shows the distribution of galaxies around the location of the centre of a typical mock catalogue. Each galaxy is weighted with a Gaussian radial distribution function $f(r)=e^{-r^2/2R^2}$ (here $R=50$ \hmpc). The radial distribution is shown in the right-hand panel. The MV formalism estimates the bulk flow of this Gaussian-weighted box, by only using the mock catalogues of the kind shown in Figs~\ref{fig:DEEP} and \ref{fig:COMPOSITE} (bottom rows).
    }}
    \label{fig:nbody}
  \end{flushleft}
\end{figure*}

\subsection{Catalogues}
\label{sec:CATS}

We create mocks of three different peculiar velocity surveys from the simulations:   i) The `DEEP' compilation includes 103 SNIa \citep{TonSchBar03}, 70 Spiral Galaxy Culsters (SC) Tully-Fisher (TF) clusters \citep{GioHaySal98, DalGioHay99}, 56  Streaming Motions of Abell Clusters (SMAC) fundamental plane (FP) clusters \citep{HudSmiLuc99, HudSmiLuc04}, 50 Early-type Far Galaxies (EFAR) FP clusters \citep{ColSagBur01} and 15 TF clusters  \citep{Wil99b}. The DEEP catalogue consists of 294 data points with a characteristic MLE depth of 50 \hmpc, calculated using $\sum w_n r_n / \sum w_n$ where the MLE weights are $w_n=1/(\sigma_n^2 + \sigma_*^2)$. In this paper, we assume $\sigma_*=150$ \kmSec. We have tried $\sigma_*=150-450$ \kmSec\ and it does not change our results appreciably. ii) The SFI++ (Spiral Field I-band) catalogue \citep{sfi1,sfi2,sfierr09} is the densest and most complete peculiar velocity survey of field spirals to date.  We use the data from the corrected dataset \citep{sfierr09}, the sample consists of 2821 TF field galaxies. The characteristic depth is 34 \hmpc. iii) The `COMPOSITE' catalogue is a compilation of the DEEP and SFI++ catalogues as well as the group SFI++ catalogue \citep{sfierr09}, the Early-type Nearby Galaxies (ENEAR; \citealt{daCBerAlo00, BerAlodaC02b, WegBerWil03}) survey and a surface brightness fluctuations (SBF) survey \citep{TonDreBla01}. With 4481 data points, the COMPOSITE catalogue has a characteristic depth of 33 \hmpc. The DEEP and SFI++ catalogues are completely independent whereas the COMPOSITE is a compilation of these and other catalogues. For further details on these catalogues see Papers I and II.

We have used these particular catalogues to investigate the effect of geometry and density on our results. The reason for using these catalogues is that we want to compare the results using a very sparse catalogue (DEEP) and the better sky coverage and higher density of the COMPOSITE catalogue. We chose the SFI++ catalogue as an intermediate case study. We tested our MV formalism on the DEEP, SFI++ and COMPOSITE mocks extracted from the LD and HR simulations. As we mentioned earlier, we extracted 4100 mocks from the LD simulations and 5000 from the HR simulation. The results based on the 5000 mock surveys from the HR simulation are virtually identical to the ones for the LD simulations. As such, in the rest of this paper, we display results only for the 4100 mocks extracted from the LD simulations. Moreover, since our results for the SFI++ catalogue are very similar to the ones for the DEEP and COMPOSITE catalogues, we do not display SFI++ results. 

In Figs~\ref{fig:DEEP} and \ref{fig:COMPOSITE}, we show the DEEP and COMPOSITE real catalogues (top rows) and a sample mock catalogue (bottom rows). The N-body simulations do not have as many close by objects as there are in the COMPOSITE catalogue, which is why the COMPOSITE mocks match the radial distribution only beyond $\twid50h^{-1} \textrm{Mpc}$.\

In Fig.~\ref{fig:nbody}, we show the weighted distribution of galaxies around the location of the centre of a typical mock catalogue (left-hand panel) and its radial distribution (right-hand panel).   Each galaxy is weighted with a Gaussian radial distribution function $f(r)=e^{-r^2/2R^2}$ with $R=50$\hmpc.   The MV formalism is designed to obtain the best estimate of the bulk flow of this Gaussian-weighted box, by only using the mock catalogues of the kind shown in Figs~\ref{fig:DEEP} and \ref{fig:COMPOSITE} (bottom rows).   Note that the Gaussian-weighted box does not have a perfect Gaussian distribution but it comes close to being one. Denser simulations would be required to test the MV formalism more rigorously.

\subsection{Mock extraction procedure}
\label{sec:Extract}

Once we have identified a random point in the N-body simulation box, we extract a set of galaxies that has the same radial selection function about this point as the catalogue we are creating mocks of.   We do not impose the additional constraint on the mocks that they must also have the same angular distribution as the real surveys for two reasons: (i) the N-body simulations are not dense enough to give us mocks that are $\it{exactly}$ like the real surveys and (ii) the weights $w_{i,n}$ of the real surveys typically depend only on the radial distribution and the velocity errors of the survey objects. Consequently, the mocks in Figs~\ref{fig:DEEP} and \ref{fig:COMPOSITE} have a relatively featureless angular distribution.   To make the mocks more realistic, we also impose a $10^\degree$ latitude zone-of-avoidance cut.

From the simulations we find the angular position, the true line-of-sight peculiar velocity $v_s$ and the redshift $cz=d_s+v_s$ for each mock galaxy, where $d_s$ is the true radial distance of the mock galaxy from the random centre we selected, all in \kmSec . We then perturb the true radial distance $d_s$ of the mock galaxy with a velocity error drawn from a Gaussian distribution of width equal to the corresponding real galaxy's velocity error, $\sigma_n$. Thus, $d_p=d_s+\delta_d$, where $d_p$ is the perturbed radial distance of the mock galaxy (in \kmSec) and $\delta_d$ is the velocity error. The mock galaxy's measured line-of-sight peculiar velocity $v_p$ is then assigned to be $v_p=cz-d_p$, where $cz$ is the redshift we found above. The reason for this procedure is that the weight we assign to each galaxy in the mock catalogues will then be similar to the weights of the real catalogues, since these depend on the radial distribution errors of the survey objects.

This procedure of perturbing the distances $d_s$ and then assigning the velocities $v_p$ to the mock galaxies introduces a Malmquist bias. We have checked the effect of the bias by following a slightly different approach to generate the mocks. We used the exact distances $d_s$ and only perturbed the velocities as $v_p=v_s+\delta_d$. We found the effect of Malmquist bias on our MV analyses to be negligible.

\begin{figure}
     \includegraphics[width=\columnwidth]{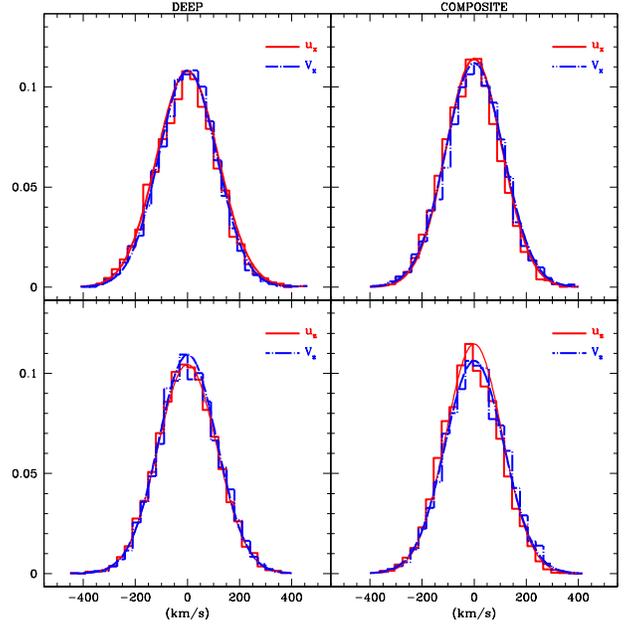}
        \caption{\small{Histograms showing the normalized probability distribution for the MV- and Gaussian-weighted bulk flow moments within a Gaussian window of radius $R=50$ \hmpc\ for the directions $x$ and $z$ in the top and bottom rows, respectively for the two types of mock catalogues in the LD simulations: DEEP (left-hand column) and COMPOSITE (right-hand column) as in Fig.~\ref{fig:LD_diff}.  The MV-weighted bulk flow moments $u_i$ are the solid histogram. The Gaussian-weighted moments $V_i$ are shown as dashed histogram.   We also superimpose a Gaussian centred at zero with width of the rms calculated. It is clear that the distributions of both the MV- and Gaussian-weighted moments are Gaussian distributed. We do not show the $y$-direction since it is statistically identical to the $x$-direction. The SFI++ catalogue shows very similar trends and so was not displayed.
        }}
     \label{fig:LD_prob}
\end{figure}

\section{Bulk Flow Moments}
\label{sec:BF}

\begin{figure}
     \includegraphics[width= \columnwidth]{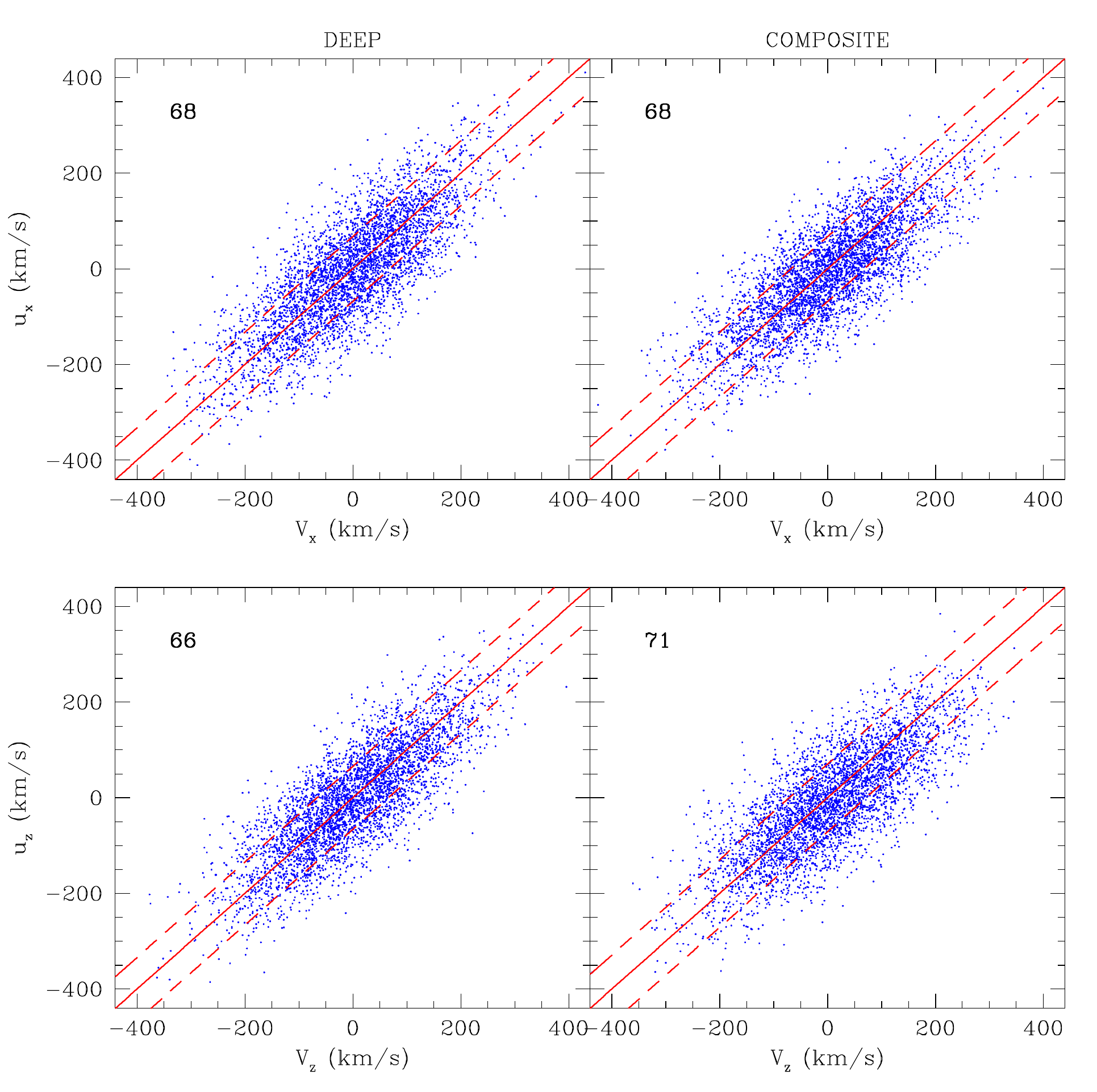}
        \caption{\small{The MV bulk flow moments $u_i$ versus the Gaussian-weighted moments $V_i$ for $R=50$ \hmpc\ for the two types of mock catalogues in the LD simulations: DEEP (left-hand column) and COMPOSITE (right-hand column). There are 4100 mocks for each of the catalogues.  We show the moments $u_x$ and $u_z$ in the top and bottom rows, respectively. The MV- and the Gaussian-weighted moments are plotted against each other (dots). A perfect correlation would put all 4100 dots on the diagonal.  The rms scatter (\kmSec) in the MV moments is displayed at the top left-hand side of each panel and shown as dashed lines. We do not show the $y$-direction since it is statistically identical to the $x$-direction. The SFI++ catalogue shows very similar trends and so was not displayed.
        }}
    \label{fig:LD_corr}
\end{figure}

\begin{figure}
     \includegraphics[width=\columnwidth]{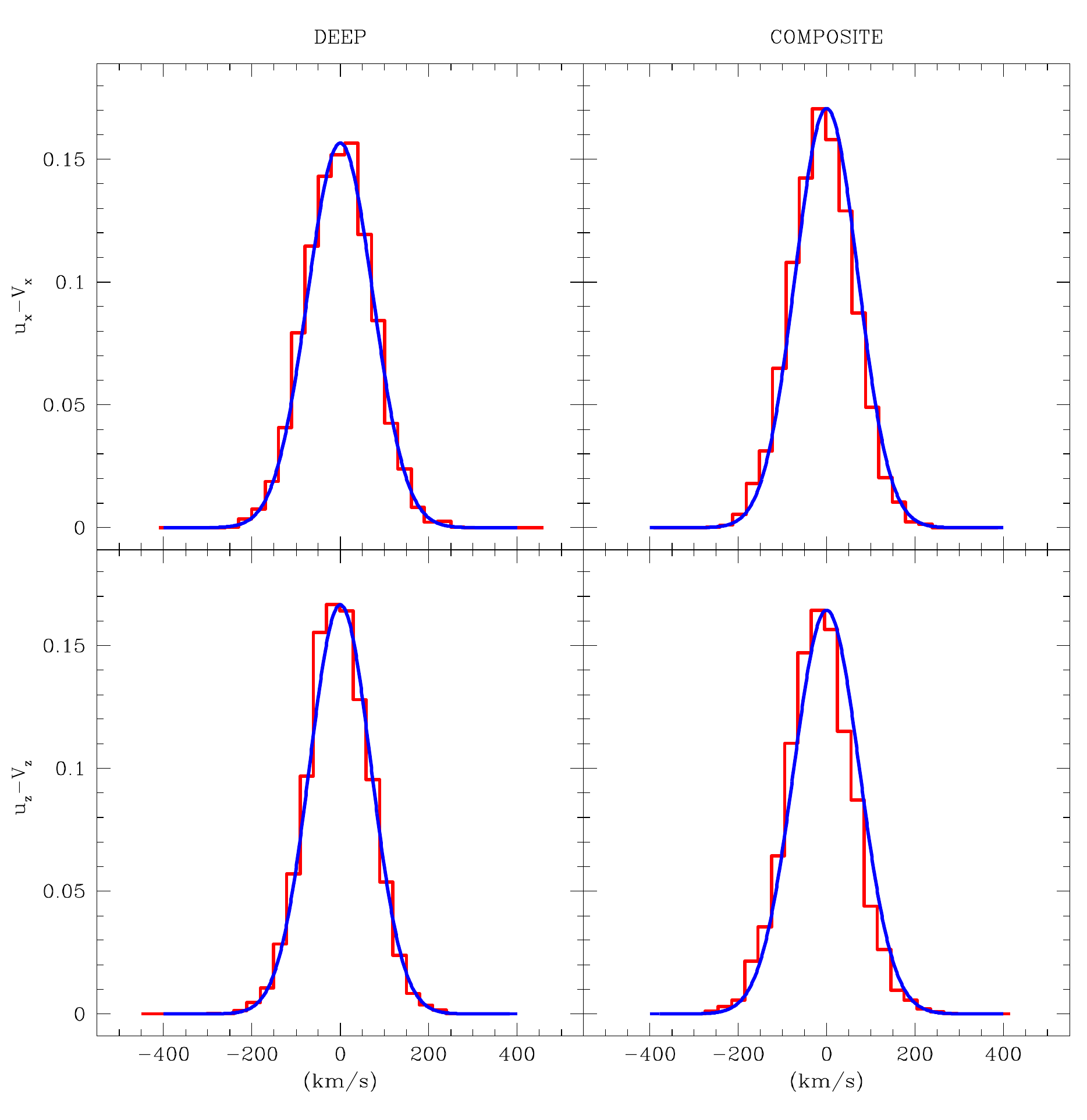}
        \caption{\small{Histograms showing the normalized probability distribution for the differences between the MV- and Gaussian-weighted moments for the $x$- and $z$- directions in the top and bottom rows, respectively.   The solid histograms show the quantities $(u_i-V_i)$ for the 4100 mock catalogues extracted from the 41 LD simulation boxes for $R=50$ \hmpc: DEEP (left-hand column) and COMPOSITE (right-hand column).   Superimposed on the histograms are Gaussians centred at zero and with the same width, $\langle (u_i-V_i)^2\rangle^{\frac12}$, as the corresponding histogram.   The fact that the distributions are centred on zero demonstrates that the MV estimators are not biased.   We do not show the $y$-direction since it is statistically identical to the $x$-direction.   The SFI++ catalogue shows very similar trends and so was not displayed.
        }}
     \label{fig:LD_diff}
\end{figure}

For each of the 4100 LD (5000 HR) mocks, we estimated the bulk flow moments  \{$u_x, u_y, u_z$\} using our MV weighting scheme (Sec.~\ref{sec:MV}).   We then compared the results to the Gaussian-weighted bulk moments \{$V_x, V_y, V_z$\} calculated by going to the same central points for each of the 4100 LD (5000 HR) mock catalogues and averaging the velocities of all the galaxies in the simulation box, each galaxy being weighted by a Gaussian weight of width $R=50$\hmpc.   Although the results that we show here are for a particular scale of $R=50$\hmpc, we have repeated our analysis for other values of $R$ with similar results.     It is worth mentioning here that since the position and the velocity of every galaxy in the N-body simulations are known exactly, their respective uncertainties are zero.   Here we present our results only from the LD simulations.   The HR simulation shows very similar results.

In Fig.~\ref{fig:LD_prob}, we show the probability distribution for the the 4100 MV-weighted bulk flow moments $u_i$ (solid) and the Gaussian-weighted moments $V_i$ (dashed) within a Gaussian window of radius $R=50$ \hmpc\  for the LD simulations.   As shown in Fig.~\ref{fig:LD_prob}, the distributions for the MV-estimated bulk flow moments (solid histogram) and the Gaussian-weighted moments (dashed histogram) are both Gaussian distributed.   This is as expected for large scale moments and reflects the fact that non-linear motions, which can lead to non-Gaussian tails in the velocity distributions for individual galaxies, have been effectively averaged out.   The widths of the distributions match well with the expectations from linear theory,

\begin{eqnarray}
 \label{eq:sigv}
 \sigma_v^2(R) &=& {H_0^2\Omega_{\rm m}^{1.1}\over{2\pi^2}}\int   dk\  P(k)W_v^2(kR),
\end{eqnarray}

where $\sigma_v(R)$ is the RMS value of the peculiar velocity field smoothed with a suitable filter with a characteristic scale $R$; $W_v(k,R)$ is the window function (Fourier transform of the filter) and $P(k)$ is the matter power spectrum.   A $\Lambda$CDM model with WMAP7 \citep{wmap7} central parameters, together with a Gaussian window function $W_v(k,R)=e^{-(kR)^2/2}$, predicts a 110 \kmSec\ width for $R=50$ \hmpc, virtually identical to the ones shown in Fig.~\ref{fig:LD_prob}.   In Paper II we estimated that for a $\Lambda$CDM model with WMAP7 central parameters, the chance of getting a $\sim400$ \kmSec\ bulk flow for a survey on scales of $50$ \hmpc\ is $\sim1$ per cent.  Examining Fig.~\ref{fig:LD_prob} confirms that the probability will be similarly small.   Indeed, the frequency of mock catalogues with $>400$ \kmSec\ was found to be comparable to the 1 per cent value.

\begin{table*}
\caption{The theoretical distribution width $\langle (u_i-U_i)^2\rangle^{\frac12}$ for the real catalogues in the first ($x$), second ($y$) and third ($z$) columns, calculated in linear theory using Eq.~\ref{eq:variance}, Eq.~\ref{eq:qi}, Eq.~\ref{eq:gmn} and Eq.~\ref{eq:fmn}. 
In the fourth ($x$), fifth ($y$) and sixth ($z$) columns, we show the widths $\langle (u_i-V_i)^2\rangle^{\frac12}$ of the $(u_i-V_i)$ histograms for the LD mocks (see Fig.~\ref{fig:LD_diff}), this should be compared to the first three columns.
The theoretical widths for the LD mocks are shown in the seventh ($x$), eighth ($y$) and ninth ($z$) columns. For the LD mocks, we quote the mean and standard deviation values of $\langle (u_i-U_i)^2\rangle^{\frac12}$, for the 4100 mock catalogues. These values are based on WMAP7 \citep{wmap7} central power spectrum parameters. 
In the last column we show the width of the distribution of the moments $u_i$ over the 4100 mock catalogues (see Fig.~\ref{fig:LD_prob}). Since the widths $u_x, u_y$ and $u_z$ were all found to be very similar, we only quote a single value for $u_i$ in the last column. All values are in \kmSec .}
\begin{tabular}{|l|c|c|c|c|c|c|c|c|c|c|}
 \hline
 & \multicolumn{3}{c}{Real catalogues} &  \multicolumn{7}{c}{LD mock catalogues} \\
 & \multicolumn{3}{c}{$\langle (u_i-U_i)^2\rangle^{\frac12}$} &  \multicolumn{3}{c}{$\langle (u_i-V_i)^2\rangle^{\frac12}$} & \multicolumn{3}{c}{$\langle (u_i-U_i)^2\rangle^{\frac12}$} & Width \\ \hline
DEEP  		& 73.28 	& 80.27 & 54.21 & 73.79 & 74.95 & 69.62 & 65.16  $\pm$ 2.71  & 65.37  $\pm$  2.79 &  57.32  $\pm$  1.98  & 111 \\ 
SFI++  		& 57.37 	& 58.63 & 47.42 & 72.38 & 73.48 & 72.60 & 56.52  $\pm$ 2.53  & 56.96  $\pm$  2.58 &  51.37  $\pm$  2.20  & 111 \\ 
COMPOSITE  	& 46.75	& 47.34 & 34.93 & 70.69 & 71.47 & 71.34 & 42.16  $\pm$ 2.15  & 42.59  $\pm$  1.60 &  39.87  $\pm$  1.17  & 112 \\ \hline 
\end{tabular}
\label{tab:variance}
\end{table*}

In Fig.~\ref{fig:LD_corr}, we show the bulk flow moments in the $x$- and $z$- directions (in the top and bottom rows, respectively) for the 4100 DEEP (left-hand column) and COMPOSITE (right-hand column) mock catalogues, extracted from the 41 LD simulation boxes.   The MV-weighted moments $u_i$ and the corresponding Gaussian-weighted moments $V_i$ are plotted against each other (dots) and the positive correlation between the two is clearly visible.   A perfect correlation would put all 4100 points on the diagonal.   

In Fig.~\ref{fig:LD_diff}, we show the probability distribution for the difference between the MV-weighted bulk flow moments $u_i$ and the Gaussian-weighted ideal moments $V_i$ for the 4100 mock surveys from the LD simulations.   A Gaussian centred at zero and with the same width as the probability distribution is also shown.   The fact that the distributions are centred on zero demonstrates that the MV estimators are not biased.

Given a mock catalogue, the theoretical expectation value for the width of the distribution, \ie $\langle (u_i-U_i)^2\rangle^{\frac12}$, can be calculated in linear theory using Eq.~\ref{eq:variance}, Eq.~\ref{eq:qi}, Eq.~\ref{eq:gmn} and Eq.~\ref{eq:fmn}.   To check the robustness of our MV method, this can then be compared with the distribution width $\langle (u_i-V_i)^2\rangle^{\frac12}$ calculated directly from the simulations [the $(u_i-V_i)$ distribution is shown in Fig.~\ref{fig:LD_diff}] using the same cosmological model.   The theoretical widths $\langle (u_i-U_i)^2\rangle^{\frac12}$ for the 4100 LD mocks are shown in Table~\ref{tab:variance}, columns 7 -- 9.   Since each mock catalogue has in principle a slightly different expectation for the width, we quote the average and standard deviation of the widths obtained from the set of the mock catalogues.   The widths $\langle (u_i-V_i)^2\rangle^{\frac12}$ found in the simulations are shown in Table~\ref{tab:variance}, columns 4 -- 6.

\begin{figure*}
  \begin{flushleft}
   \centering
    \begin{minipage}[c]{1.0\textwidth}
      \centering
      \includegraphics[width=8.cm]{nbody1.pdf}
      \includegraphics[width=8.cm]{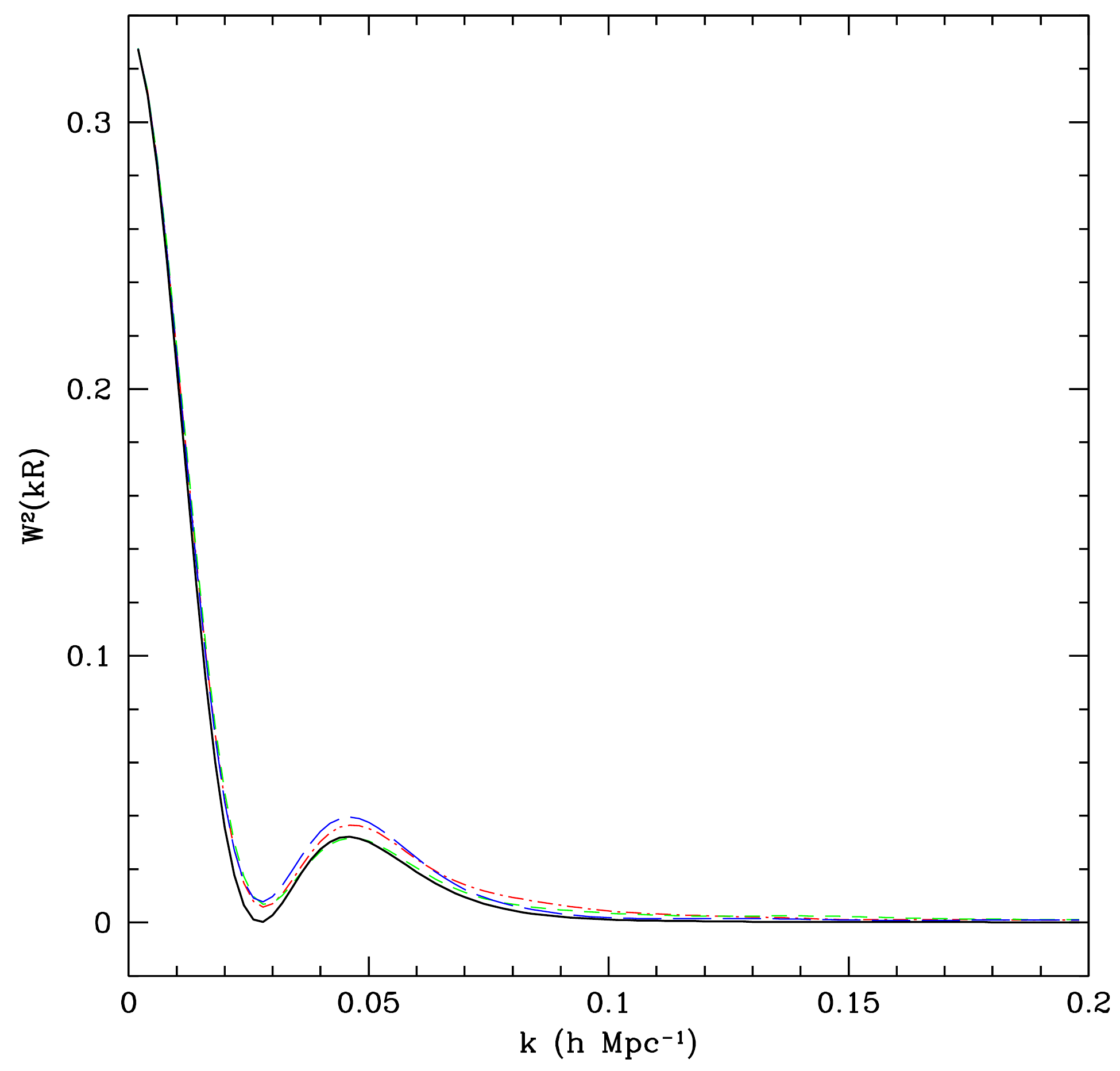}
    \end{minipage}
    \caption{\small{\textit{Left}: the distribution of galaxies around the location of the centre of a typical mock catalogue.   Each galaxy is weighted with a Gaussian radial distribution function $f(r)=e^{-r^2/2R^2}$ (here $R=50$ \hmpc).   \textit{Right}: The window functions $W^2_{ii}$ (see Eq.~\ref{eq:Wij}) of the bulk flow components $u_i$, for $R=50$ \hmpc.   The $x, y$ and $z$ components are dash-dotted, short-dashed, long-dashed lines, respectively, and correspond to the distribution in the left-hand panel.   The solid line is the ideal window function (since the ideal survey is isotropic, all components are the same).
    }}
    \label{fig:nbody_wf}
  \end{flushleft}
\end{figure*}

Comparing linear theory predictions [$\langle (u_i-U_i)^2\rangle^{\frac12}$ in Table~\ref{tab:variance}, columns 7 -- 9] with the widths found in the numerical simulations [$\langle (u_i-V_i)^2\rangle^{\frac12}$, columns 4 -- 6], we see that the distribution width found in the simulations are somewhat different than the widths predicted by linear theory.   This is due to the fact that the simulations are not dense enough and thus do not have enough galaxies to emulate an ideal survey.   We explain this through Fig.~\ref{fig:nbody_wf}.   In the left-hand panel, we show the weighted distribution of galaxies around the location of the centre of a typical mock catalogue.   Each galaxy is weighted with a Gaussian radial distribution function $f(r)=e^{-r^2/2R^2}$ with $R=50$\hmpc.   The right-hand panel shows the window functions $W^2_{ii}$ of the bulk flow components $u_i$ for this distribution (dash-dotted, short-dashed and long-dashed lines for the $x,y$ and $z$ components, respectively) and the ideal window function (solid line).   Non-Gaussianity in the distribution of galaxies in the left-hand panel causes a slight mismatch between its window functions and the ideal one.   With a larger number of galaxies in the simulations, the Gaussian-weighted moments $V_i$ would approach the ideal moments $U_i$, and give a closer match between $\langle (u_i-U_i)^2\rangle^{\frac12}$ and $\langle (u_i-V_i)^2\rangle^{\frac12}$.   The DEEP catalogue does not have as many close by galaxies as in the SFI++ and COMPOSITE catalogues, and thus the variance estimates calculated using linear theory [$\langle (u_i-U_i)^2\rangle^{\frac12}$] and the LD and HR simulations [$\langle (u_i-V_i)^2\rangle^{\frac12}$] are significantly closer to each other.   Taken together with the lack of bias (see Fig.~\ref{fig:LD_diff}), it is clear that non-linear motions are not having a significant effect on these large-scale moments.

\begin{figure}
     \includegraphics[width= \columnwidth]{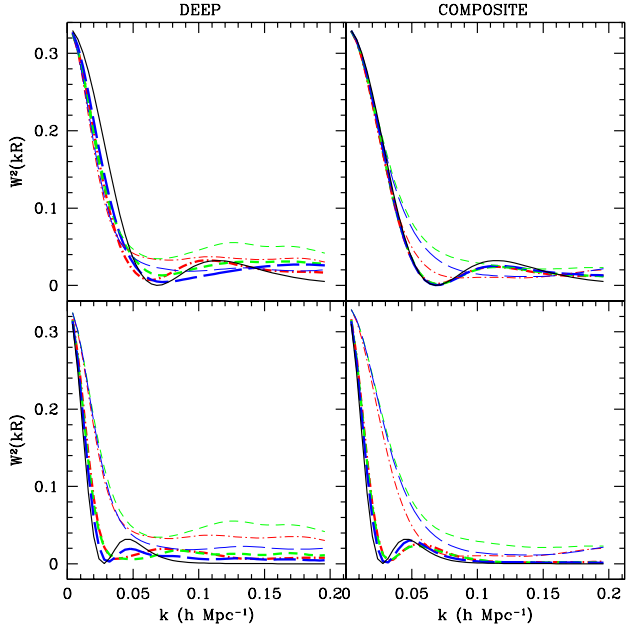}
        \caption{\small{The window functions $W^2_{ii}$ of the bulk flow components calculated using MV (thick) weights (see Eq.~\ref{eq:weight}) and MLE (thin) weights (see Eq.~\ref{eq:mle}) for $R=20$ \hmpc\ (top row) and $R=50$ \hmpc\ (bottom row) for the DEEP (left-hand column) and COMPOSITE (right-hand column) real catalogues.   The $x, y$ and $z$ components are dash-dotted, short-dashed, long-dashed lines, respectively.   The solid line is the ideal window function.
        }}
    \label{fig:non_linearity}
\end{figure}

The much improved performance of MV formalism over the widely used MLE scheme is also evident in Fig.~\ref{fig:non_linearity}, where we show the window functions $W^2_{ii}$ of the bulk flow components, calculated using MV (thick) and MLE (thin) methods.   These window functions correspond to the DEEP (left-hand column) and COMPOSITE (right-hand column) real catalogues, for $R=20$ \hmpc\ (top row) and $R=50$ \hmpc\ (bottom row).   For both DEEP and COMPOSITE catalogues, the MV window functions are a reasonable match to the ideal ones.   The MLE window functions are not only contaminated by small-scale power, they are also very different for the $x$-, $y$- and $z$-directions -- making it difficult to interpret the MLE bulk flow moments. On the other hand, by directly controlling the survey window functions the MV formalism effectively suppresses the small-scale contribution to the bulk flow. Since it is the small-scales that are predominantly plagued by non-linear effects, the MV scheme is able to make a clean estimate (compared to MLE) of the bulk flow components, while keeping the non-linear contamination to a minimum.

In Table~\ref{tab:variance}, columns 1 -- 3, we also show the values of the theoretical widths $\langle (u_i-U_i)^2\rangle^{\frac12}$ from the real catalogues on which the mocks are based.   We see that the theoretical widths for the real catalogues (columns 1 -- 3) are somewhat larger than the theoretical widths for the mocks (columns 7 -- 9). This is due primarily to the fact that the objects in the simulated catalogues are less clumped than in the real catalogues, even though they have similar radial distribution functions.   This is evident in Figs~\ref{fig:DEEP} and \ref{fig:COMPOSITE}, where the mock catalogues can be seen as having a relatively featureless spatial distribution. Less clumping and fewer close by galaxies in the simulations lower the MV-weighted bulk flow moments $u_i$, resulting in somewhat lower widths $\langle (u_i-U_i)^2\rangle^{\frac12}$ than the real catalogue widths. The creation of mock catalogues with widths that more closely matched the real catalogue widths would require simulations with higher resolution.

We also found that the sparser the mock catalogue is (eg. DEEP), the higher the chances of getting large velocities (see the extended tails in the velocity distributions for the DEEP mocks in Fig.~\ref{fig:LD_prob}), but in a way that is consistent with the larger uncertainties associated with the estimators derived from these mock catalogues.  This can be seen by comparing the predicted distribution widths $\langle (u_i-U_i)^2\rangle^{\frac12}$ for the DEEP and COMPOSITE mock catalogues in Table~\ref{tab:variance}, columns 7 -- 9. The DEEP mocks, being sparser compared to the COMPOSITE mocks, have larger widths.   Comparing the widths of $(u_i-V_i)$ histograms  (Table~\ref{tab:variance}, columns 4 -- 6) found in the simulations (Fig.~\ref{fig:LD_diff}), we again see that the DEEP mocks have marginally  larger uncertainties in the bulk estimators, as expected.

\section{Discussion and conclusions}
\label{sec:Conclude}

In previous papers (Papers I and II), we developed a weighting scheme for analyzing peculiar velocity surveys that gives estimators of idealized bulk flow moments that reflect the flow of a volume of a particular scale centred on our location rather than the characteristics of a particular survey.   Given a peculiar velocity survey, the MV method is capable of `redesigning' the survey window function in a way that minimizes the aliasing of small-scale power on to large scales, thereby making comparisons with linear theory as well as among independent surveys possible. The direct control over a survey window function makes the MV formalism an extremely useful tool when comparing bulk flow results across independent surveys with varying characteristics.

Using mock catalogues drawn from numerical simulations, we have demonstrated that the MV formalism, within errors, recovers the bulk flow moments of the underlying matter distribution and that the MV moments are unbiased estimators of the bulk flow of a volume of a given scale, regardless of the geometry of a particular survey.   The MV moments are unbiased, in that on average they give the correct values for the idealized bulk flow components.   We calculated the variance of the bulk estimator using (i) linear theory $\langle (u_i-U_i)^2\rangle^{\frac12}$ and (ii) numerical simulations $\langle (u_i-V_i)^2\rangle^{\frac12}$.   Although the variance calculated using the simulations were found to be somewhat different from the linear theory predictions, we argued that this is due to the simulations being underdense and thus not having enough galaxies. For numerical simulations with higher resolution (more galaxies), we expect the Gaussian-weighted moments $V_i$ to approach the ideal moments $U_i$ and give a much closer match. We found the variance estimates using simulations and linear theory to be significantly closer to each other for the DEEP catalogue, which has fewer close by galaxies and thus performed much better than the SFI++ and COMPOSITE catalogues when testing the MV formalism.   These results validate our use of linear theory in the development of the MV method and confirms the fact that non-linear, small-scale motions do not significantly affect the MV estimators.

We tested many facets of the MV formalism and found agreement in all the tests we performed using the LD and HR simulations.  We found that the chance of getting large flows ($\sim400$ \kmSec) in a $\Lambda$CDM universe is of order of $\sim1$ per cent. The bulk moments $u_i$ estimated using our MV formalism are, within errors, the same as the moments $V_i$ of the volume as traced by all the galaxies in the simulation box and linear theory correctly predicts the variance of the estimators. Further, since the formalism allows for exploration of all scales where there are data, we can reliably explore flows on many scales and track the dynamics of volumes of different scales (parametrized by a radius of a Gaussian sphere $R$).

\section{Acknowledgments}
We would like to thank Mike Hudson for many thoughtful and useful comments. We are also grateful to R\'oman Scoccimarro and the LasDamas collaboration and Changbom Park and the Horizon Run collaboration for providing us with the simulations. This work was supported by the National Science Foundation through TeraGrid resources provided by the NCSA.

\bibliographystyle{mn2e}
\bibliography{MV}

\end{document}